\documentstyle[twocolumn,pra,aps]{revtex}
\def\begmc{}
\def\endmc{}

\begin{document}
\title{Friction and inertia for a mirror in a thermal field}
\author{Marc Thierry Jaekel $^{(a)}$ and Serge Reynaud $^{(b)}$}
\address{(a) Laboratoire de Physique Th\'{e}orique de l'ENS
\thanks{%
Unit\'e propre du Centre National de la Recherche Scientifique,
associ\'ee \`a l'Ecole Normale Sup\'erieure et \`a l'Universit\'e
Paris-Sud}, 24 rue Lhomond, F75231 Paris Cedex 05 France\\
(b) Laboratoire de Spectroscopie Hertzienne
\thanks{%
Unit\'e de l'Ecole Normale Sup\'erieure et de l'Universit\'e
Pierre et Marie Curie, associ\'ee au Centre National de la Recherche
Scientifique}, 4 place Jussieu, case 74, F75252 Paris Cedex 05 France}
\date{{\sc Physics Letters} {\bf A 172} (1993) 319-324}
\maketitle

\begin{abstract}
The force experienced by a mirror moving in vacuum vanishes in the case of
uniform velocity or uniform acceleration, as a consequence of spatial
symmetries of vacuum. These symmetries do not subsist in a thermal field. We
give a general expression of the corresponding viscosity coefficient valid
at any temperature and for any reflectivity function. We show that the
computed motional force also contains a non vanishing inertial term. The
associated mass correction goes to zero in the limiting cases of perfect
reflection or of zero temperature.
\end{abstract}

\begmc

\section*{Introduction}

Objects which scatter a quantum field are submitted to a radiation pressure.
When two scatterers are present in vacuum, a mean force, the so-called
Casimir force \cite{Friction1}, results for each of them. A recent
discussion and references may be found in \cite{Friction2}. For a scatterer
alone in vacuum, the force of zero mean value still has quantum fluctuations 
\cite{Friction3}, which are associated through fluctuation-dissipation
relations with a motional force \cite{Friction4}.

Using the techniques of quantum field theory, Fulling and Davies have
computed the motional force for a perfectly reflecting mirror in the vacuum
state of a scalar field in a two-dimensional spacetime. A linear
approximation (first order expansion in the mirror's displacement) leads
from their result \cite{Friction5} to a force proportional to the third time
derivative of the mirror's position 
\begin{equation}
\delta F_0 (t)=\frac{\hbar }{6\pi c^2 }\delta q^{\prime \prime \prime }(t)
\eqnum{1a}
\end{equation}
This force vanishes for a uniform velocity, as well as for a uniform
acceleration. These properties are related to spatial symmetries of the
vacuum: vacuum fields are invariant under the action of Lorentz boosts \cite
{Friction6} while they appear to a uniformly accelerating observer as
thermal fields in its own frame \cite{Friction7}. Then, no friction force,
proportional to the velocity, nor inertial force, proportional to the
acceleration, appear in the radiative reaction of vacuum fields upon the
moving mirror.

In a thermal field, the motional force becomes \cite{Friction8} 
\begin{eqnarray}
\delta F_T (t) &=&\delta F_0 (t)-\lambda _T \delta q^\prime (t) 
\eqnum{1b} \\
\lambda _T  &=&\frac{2\pi T^2 }{3\hbar c^2 }  \eqnum{1c}
\end{eqnarray}
The temperature is measured as an energy ($k_{B}=1$). As expected from the
analysis by Einstein of the Brownian motion of a scatterer in a thermal
field \cite{Friction9}, a viscous force now arises (a thermal field is not
Lorentz invariant). The viscosity coefficient $\lambda _T $ goes to zero
like $T^2 $ near the vacuum state. Still no inertial force appears in this
expression.

In a more elaborate treatment, the mirror is described by reflection and
transmission amplitudes obeying unitarity and causality requirements \cite
{Friction2}, and the mirror is supposed to become transparent above a
reflection cutoff $\omega _C $, much smaller in reduced units than the
mirror's mass $m$ 
\begin{equation}
\hbar \omega _C \ll mc^2   \eqnum2 
\end{equation}
It is then possible (neglecting the recoil effect) to compute the
susceptibility function describing the motional force in a linear
approximation \cite{Friction4}. In contrast with the unphysical model of a
perfect mirror, the susceptibility now obeys the expected causality and
stability properties \cite{Friction10,Friction8}. The associated dispersion
relations set limits on the possible values of the mirror's mass $m$. We
will not consider this problem in the present letter, although it will
certainly have to be included in a complete treatment of inertia corrections.

Here, we shall show that the expression computed for the motional
susceptibility corresponds to a non vanishing inertial force for a partially
transmitting mirror scattering a thermal field. In conformity with the
particular results stated above, the associated mass correction goes to zero
in the limiting cases of perfect reflection or zero temperature.

\section*{Linear response theory at the quasistatic limit}

In a first order expansion in the mirror's displacement $\delta q$ (linear
approximation), the motional force $\delta F_T $ can be written in terms of
a susceptibility $\chi _T $ 
\begin{eqnarray}
&&\delta F_T (t)={\int }{\rm d}\tau \chi _T (\tau )\delta q(t-\tau ) 
\eqnum{3a} \\
&&\delta F_T [\omega ]=\chi _T [\omega ]\delta q[\omega ]  \eqnum{3b}
\end{eqnarray}
where we denote for any function $f$ 
\[
f(t)={\int }\frac{{\rm d}\omega }{2\pi }f[\omega ]e^{-i\omega t} 
\]
In contrast with the usual models of quantum Brownian motion lying upon a
linear coupling between mirror and fields, the mirror is here coupled to a
quantity, the radiation pressure, quadratic in the fields. Then, the
susceptibility depends upon temperature and can be written in terms of the
reflection and transmission amplitudes $r$ and $s$ as \cite{Friction8} 
\begin{eqnarray}
\chi _T [\omega ] &=&\frac{i\hbar }{2c^2 }{\int }\frac{{\rm d}\omega
^\prime }{2\pi }\omega ^\prime (\omega -\omega ^\prime )\alpha [\omega
^\prime ,\omega -\omega ^\prime ]  \nonumber \\
&&\times \left( \varepsilon _T [\omega ^\prime ]+\varepsilon _T [\omega
-\omega ^\prime ]\right)  \eqnum{4a} \\
\alpha [\omega ,\omega ^\prime ] &=&1+r[\omega ]r[\omega ^{\prime
}]-s[\omega ]s[\omega ^\prime ]  \eqnum{4b} \\
\varepsilon _T [\omega ] &=&\coth \frac{\hbar \omega }{2T}  \eqnum{4c}
\end{eqnarray}
It can also be written ($\widetilde{\xi }_T $ and $\xi _T $ real functions
of $\omega $) 
\[
\chi _T [\omega ]=\widetilde{\xi }_T [\omega ]+i\xi _T [\omega ] 
\]
The dissipative part $\xi _T $ is the commutator of the radiation pressure
force operator $F$, and is related to the stationary correlation function 
$C_T $ computed for a motionless mirror \cite{Friction4,Friction8} 
\begin{eqnarray*}
\xi _T (t) &=&\frac{\left\langle \left[ F(t),F(0)\right] \right\rangle }
{2\hbar }=\frac{C_T (t)-C_T (-t)}{2\hbar } \\
C_T (t) &=&\left\langle F(t)F(0)\right\rangle -\left\langle F\right\rangle
^2 
\end{eqnarray*}
The dispersive part $\widetilde{\xi }_T $ can be deduced from the
dissipative one $\xi _T $ through dispersion relations \cite
{Friction10,Friction8}.

In order to give a precise evaluation of the inertia corrections, we now
introduce a quasistatic expansion of the motional force (3) 
\begin{eqnarray}
&&\delta F_T (t)=-\left( \lambda _T \delta q^\prime (t)+\mu _T \delta
q^{\prime \prime }(t)+\ldots \right)  \eqnum{5a} \\
&&\chi _T [\omega ]=i\omega \lambda _T +\omega ^2 \mu _T +\ldots 
\eqnum{5b} \\
&&\lambda _T =-i\chi _T ^\prime [0]\qquad \mu _T =\frac{\chi
_T ^{\prime \prime }[0]}2   \eqnum{5c}
\end{eqnarray}
$\chi _T [0]$, which would describe a position dependent static force,
vanishes since the mean radiation pressure is zero in a thermal state. The
coefficients $\lambda _T $ and $\mu _T $ are respectively associated with
viscous and inertial forces \footnote{Note that $\mu _T $ has not the same 
definition as in ref. \cite{Friction8}.}; we are not interested in higher 
order quasistatic coefficients.

The functions $\xi _T $ and $C_T $ are connected through a
fluctuation-dissipation relation \cite{Friction11,Friction8} 
\[
C_T [\omega ]=\frac{2\hbar }{1-e^{-\frac{\hbar \omega }T }}\xi _T [\omega
] 
\]
At the low frequency limit where ($\lambda _T $ and $\mu _T $ are real) 
\begin{eqnarray*}
\xi _T [\omega ]\approx \omega \lambda _T  &\qquad &{\rm for}\ \omega
\rightarrow 0 \\
C_T [\omega ]\approx \frac{2T}\omega \xi _T [\omega ]\approx 2T\lambda
_T  &\qquad &{\rm for}\ \omega \rightarrow 0
\end{eqnarray*}
Einstein's relation \cite{Friction9} between the viscosity coefficient 
$\lambda _T $ and the momentum diffusion coefficient is recovered $\left( 
\frac{1}2 C_T [0]=T\lambda _T \right) $, whatever the expression of 
$\lambda _T $ may be (one has only supposed $\lambda _T \neq 0$).

\section*{Evaluation of the quasistatic coefficients}

We now evaluate the two coefficients $\lambda _T $ and $\mu _T $, first in
the vacuum state, then in a thermal state.

At zero temperature, the function $\varepsilon _T $ appearing in equations
(4) coincides with the sign function ($\varepsilon (\omega )=\frac\omega 
{\left| \omega \right| }$), and the known result \cite{Friction4} is
recovered 
\[
\chi _0 [\omega ]=\frac{i\hbar }{c^2 }{\int_0 ^\omega }\frac{{\rm d}
\omega ^\prime }{2\pi }\omega ^\prime (\omega -\omega ^\prime )\alpha
[\omega ^\prime ,\omega -\omega ^\prime ] 
\]
At a non zero temperature, we write the `smoothed' sign function $
\varepsilon _T $ as the sum of the sign function $\varepsilon $ and of a
correction involving the mean number $n_T $ of thermal photons per mode 
\begin{eqnarray*}
\varepsilon _T [\omega ] &=&\varepsilon (\omega )+\delta \varepsilon
_T [\omega ] \\
\delta \varepsilon _T [\omega ] &=&2\varepsilon (\omega )n_T \left[ \left|
\omega \right| \right] \\
n_T [\omega ] &=&\frac{1}{e^{\frac{\hbar \omega }T }-1}
\end{eqnarray*}
We then deduce 
\begin{eqnarray*}
\chi _T [\omega ] &=&\chi _0 [\omega ]+\delta \chi _T [\omega ] \\
\delta \chi _T [\omega ] &=&\frac{2i\hbar }{c^2 }{\int_0 ^\infty }\frac{
{\rm d}\omega ^\prime }{2\pi }\omega ^\prime n_T [\omega ^\prime ] \\
&\times& \left( (\omega -\omega ^\prime )\alpha [\omega ^\prime ,\omega
-\omega ^\prime ] \right. \\
&&\left. +(\omega +\omega ^\prime )\alpha [-\omega ^{\prime
},\omega +\omega ^\prime ]\right)
\end{eqnarray*}
The function $\chi _0 [\omega ]$ scales as $\omega ^{3}$ at low frequencies
so that the coefficients $\chi _0 ^\prime [0]$ and $\chi _0 ^{\prime
\prime }[0]$ vanish. A straighforward calculation thus leads to the
following expression of the viscosity coefficient $\lambda _T $ defined by
equations (5) 
\begin{eqnarray}
\lambda _T  &=&\frac{2\hbar }{c^2 }{\int_0 ^\infty }\frac{{\rm d}\omega 
}{2\pi }n_T [\omega ]\partial _\omega \left( \omega ^2 a[\omega ]\right)
\eqnum{6a} \\
a[\omega ] &=&1+r[\omega ]r[-\omega ]-s[\omega ]s[-\omega ]  \eqnum{6b}
\end{eqnarray}
This expression may be transformed, first by integrating by parts, then by
using the fact that $n_T [\omega ]$ is a function of $\frac\omega T $
only 
\[
-\omega \partial _\omega n_T [\omega ]=T\partial _T n_T [\omega ] 
\]
One eventually obtains 
\begin{eqnarray}
\lambda _T  &=&\frac{2T}{c^2 }\frac{{\rm d}A}{{\rm d}T}  \eqnum{6c} \\
A(T) &=&{\int_0 ^\infty }\frac{{\rm d}\omega }{2\pi }\hbar \omega
n_T [\omega ]a[\omega ]  \eqnum{6d}
\end{eqnarray}
In a similar manner, we get the mass correction $\mu _T $ defined by
equations (5) 
\begin{eqnarray}
\mu _T  &=&\frac{\hbar }{c^2 }{\int_0 ^\infty }\frac{{\rm d}\omega }
{2\pi }n_T [\omega ]\partial _\omega \left( \omega ^2 b[\omega ]\right) 
\eqnum{7a} \\
b[\omega ] &=&i\left( r^\prime [\omega ]r[-\omega ]+r[\omega ]r^{\prime
}[-\omega ]\right)  \nonumber \\
&&-i\left( s^\prime [\omega ]s[-\omega ]+s[\omega ]s^\prime [-\omega
]\right)  \eqnum{7b}
\end{eqnarray}
that is 
\begin{eqnarray}
\mu _T  &=&\frac T {c^2 }\frac{{\rm d}B}{{\rm d}T}  \eqnum{7c} \\
B(T) &=&{\int_0 ^\infty }\frac{{\rm d}\omega }{2\pi }\hbar \omega
n_T [\omega ]b[\omega ]  \eqnum{7d}
\end{eqnarray}
At this point, we want to emphasize that the two coefficients $\lambda _T $
and $\mu _T $ result from the radiative reaction upon the mirror of the
thermal fields only. There is no viscous nor inertial force for a partially
transmitting mirror in vacuum, as well as for a perfect one, in agreement
with the spatial symmetries discussed in the introduction. It is worth
noting that the vacuum fields may be reintroduced in expressions (6) and (7)
by replacing $n_T [\omega ]$ by $\left( \frac{1}2 +n_T [\omega ]\right) $
without changing the resulting values in equations (6a) and (7a), where the
vacuum contribution $\frac{1}2 $ leads to a null integral. Alternatively,
the term $\frac{1}2 $ contributes to equations (6d) and (7d), but its
contributions, being temperature independent, do not affect the resulting
values of entropy-like expressions (6c) and (7c). The foregoing discussion
shows that the expressions of $\lambda _T $ and $\mu _T $ could as well
have been obtained from a description of vacuum fluctuations as `zero-point
fields' \footnote{%
This description, initiated by Planck (1911), Einstein and Stern (1913),
Nernst (1916), is reviewed for example in ref. \cite{Friction12}; it
constitutes the basis of stochastic electrodynamics, reviewed in ref. \cite
{Friction12}.}. This is not true for the complete expression of the
susceptibility: when replacing $n_T [\omega ]$ by $\frac{1}2 $ in the
thermal contribution $\delta \chi _T [\omega ]$ to susceptibility, one
obtains an expression which differs from the vacuum contribution $\chi
_0 [\omega ]$. The quasistatic coefficients of higher order, especially the
coefficient $\chi ^{\prime \prime \prime }[0]$ which describes the radiative
reaction of vacuum, are not correctly obtained without fully accounting for
the quantum character of vacuum fluctuations. The difference between a
quantum and a classical description of vacuum radiation pressure is also
discussed in ref. \cite{Friction14}.

\section*{Expression in terms of the reflection probabilities and phase
shifts}

In order to interpret the expressions obtained for $\lambda _T $ and $\mu
_T $, we introduce the modulus and phase of the scattering coefficients.
Using the unitarity of the scattering matrix 
\begin{eqnarray*}
&&\left| s[\omega ]\right| ^2 +\left| r[\omega ]\right| ^2 =1 \\
&&s[\omega ]r[\omega ]^{*}+r[\omega ]s[\omega ]^{*}=0
\end{eqnarray*}
and the reality of the scattering functions (written in the time domain) 
\[
s[-\omega ]=s[\omega ]^{*}\qquad r[-\omega ]=r[\omega ]^{*} 
\]
one deduces that these modulus and phases may be written in terms of only
two functions 
\begin{eqnarray*}
\left| r[\omega ]\right| ^2  &=&R[\omega ]{\qquad }\left| s[\omega ]\right|
^2 =1-R[\omega ] \\
\frac{s[\omega ]}{s[\omega ]^{*}} &=&-\frac{r[\omega ]}{r[\omega ]^{*}}
=e^{i\Delta [\omega ]}
\end{eqnarray*}
The function $R$ is the reflection probability while $\Delta $ is the sum of
the two phase shifts associated with the scattering matrix: $e^{i\Delta
[\omega ]}$ is precisely the determinant $s[\omega ]^2 $-$r[\omega ]^2 $
of the scattering matrix. We will also define the scattering delays for
fields around frequency $\omega $; more precisely, $2\tau $ will be the sum
of the two time delays associated with the scattering matrix 
\[
2\tau [\omega ]=\Delta ^\prime [\omega ] 
\]
It turns out that the viscosity coefficient $\lambda _T $ depends only upon
the reflection probability 
\begin{equation}
a[\omega ]=2R[\omega ]  \eqnum{6e}
\end{equation}
whereas the mass correction $\mu _T $ also depends upon the phase shifts 
\begin{equation}
b[\omega ]=2\left( 1-2R[\omega ]\right) \tau [\omega ]  \eqnum{7e}
\end{equation}
In particular, the behaviour at low frequencies of $a$ and $b$, which will
play a dominant role at the low temperature limit, is determined by the
parameters $R_0 =R[\omega \rightarrow 0]$ and $\tau _0 =\tau [\omega
\rightarrow 0]$.

A simple model fulfilling the requirements of unitarity, causality and high
frequency transparency corresponds to the lorentzian scattering functions 
\begin{eqnarray*}
r[\omega ] &=&\frac{-1}{1-i\omega \tau _0 }\qquad s[\omega ]=\frac{-i\omega
\tau _0 }{1-i\omega \tau _0 } \\
R[\omega ] &=&\frac{1}{1+\omega ^2 \tau _0 ^2 }\qquad \tau [\omega ]=
\frac{\tau _0 }{1+\omega ^2 \tau _0 ^2 }
\end{eqnarray*}
The parameter $\tau _0 $, defined as the time delay evaluated at
frequencies lower than the reflection cutoff of the mirror, also appears as
the inverse of this cutoff.

\section*{Discussion of the viscosity coefficient}

Collecting equations (6), we are now able to give a simple interpretation of
the quantity 
\[
A(T)={\int_0 ^\infty }\frac{{\rm d}\omega }{2\pi }2\hbar \omega
n_T [\omega ]R[\omega ] 
\]
This is indeed the energy flux associated with the two thermal fields coming
onto the mirror from the left and from the right sides ($\frac{{\rm d}\omega 
}{2\pi }\hbar \omega n_T [\omega ]$ is the energy flux in the band ${\rm d}
\omega $ for one propagation direction and the factor $2$ stands for the two
input fields) integrated over the reflection bandwidth $R[\omega ]$ of the
mirror. It is easily shown that $A(T)$ increases with $T$, so that $\lambda
_T $ is positive and corresponds effectively to a damping force. Simple
expressions are obtained at the low and high temperature limits.

At the high temperature limit ($T\gg \hbar \omega _C $ where $\omega _C $
is the reflection cutoff), $n_T [\omega ]$ can be replaced by its classical
approximation ($\hbar \omega n_T [\omega ]\approx T$) and the incident
energy flux is the product of the temperature by the bandwidth 
\[
A(T)=2T\Omega _C \qquad \Omega _C ={\int_0 ^\infty }\frac{{\rm d}\omega 
}{2\pi }R[\omega ] 
\]
$\Omega _C $ is of the order of the reflection bandwidth; $\Omega _C =
\frac{\omega _C }{4}$ for the lorentzian model. As $A(T)$ is a linear
function of $T$, it follows from equation (6a) that 
\[
\lambda _T =\frac{2A(T)}{c^2 } 
\]
This result can also be understood in a simple manner. For each photon of
energy $\hbar \omega $ reflected by the mirror moving with a uniform
velocity $v$, there is a net momentum transfer $\frac{-2\hbar \omega v}{c^2}
$ to the mirror, as a consequence of the Doppler effect. The viscosity
coefficient $\lambda _T $ is therefore $\frac{2A}{c^2 }$, where $A$ is the
energy of all photons impinging on the mirror per unit time.

At the low temperature limit $T\ll \hbar \omega _C $, the reflection
probability can be replaced by its low-frequency value $R_0 $ so that one
gets $A$ in terms of the energy flux coming onto the mirror integrated over
all frequencies 
\[
A(T)=\frac{R_0 \pi T^2 }{6\hbar } 
\]
$A(T)$ is now a quadratic function of $T$, and equation (6a) implies that 
$\lambda _T $ is twice the value expected from the interpretation in terms
of Doppler shifts 
\[
\lambda _T =\frac{4A(T)}{c^2 }=\frac{R_0 2\pi T^2 }{3\hbar c^2 } 
\]
It is well known that the interpretation in terms of Doppler shifts is too
naive: taken seriously, it would lead to a viscous damping of the mirror in
vacuum. This paradox has already been elucidated \cite{Friction6}: the
Lorentz transformation affects the field amplitudes as well as the field
frequencies; when this is taken into consideration, it turns out that the
vacuum fields effectively obey Lorentz invariance and do not contribute to
damping. The results obtained in the present paper show that the
modification of the field amplitudes also plays a role at non zero
temperature. This role appears to be negligible at high temperatures, while
it amounts to double the coefficient evaluated from Doppler shifts at low
temperatures.

\section*{Discussion of the inertial force}

The expressions (7) giving the mass correction $\mu _T $ can be discussed
along the same lines. The quantity 
\[
B(T)={\int_0 ^\infty }\frac{{\rm d}\omega }{2\pi }2\hbar \omega
n_T [\omega ]\left( 1-2R[\omega ]\right) \tau [\omega ] 
\]
is an integral over frequency of the thermal energy flux, with a weight
function proportional to the time delays.

For a perfect mirror, the phase shifts do not depend upon frequency, so that
the mass correction vanishes ($\tau [\omega ]=0$) at any temperature, in
consistency with the results stated in the introduction (see eqs 1). For a
partially transmitting mirror however, the phase shifts are frequency
dependent in order to obey causality and high frequency transparency and the
scattering delays do not vanish. In a two-mirror configuration, the Casimir
energy can be expressed in terms of the phase shifts \cite{Friction2}; it
appears as a finite part of the incident field energy stocked because of the
time delays. The quantity $B$ would have the same interpretation of an
incident energy stocked because of the time delays for a mirror having a
small reflexion probability at all frequencies: the function $\left(
1-2R[\omega ]\right) $ could be replaced by 1 in its expression. This is not
always the case, since the function $\left( 1-2R[\omega ]\right) $ can even
change its sign with frequency. This prevents to give a simple
interpretation of $B$ as a stocked energy.

At the high temperature limit, one gets 
\begin{eqnarray*}
&&B(T)=T\Delta _{S} \\
&&\Delta _{S}={\int_0 ^\infty }\frac{{\rm d}\omega }{2\pi }\left(
1-2R[\omega ]\right) 2\tau [\omega ] \\
&&\mu _T =\frac T {c^2 }\frac{{\rm d}B}{{\rm d}T}=\frac{B(T)}{c^2 }
\end{eqnarray*}
For instance, the model of lorentzian scattering coefficients leads to the
value $B=0$. This differs from the expression $\frac T 2 $ of the stocked
energy computed from the same expressions with $\left( 1-2R[\omega ]\right) $
replaced by 1.

At the low temperature limit, the functions $R$ and $\tau $ can be replaced
by their low frequency values 
\[
B(T)=\left( \frac{1}{R_0 }-2\right) \tau _0 A(T)=\left( 1-2R_0 \right) 
\frac{\tau _0 \pi T^2 }{6\hbar }
\]
$A$ is the integrated energy flux, discussed in the previous section. One
deduces 
\[
\mu _T =\frac{2B(T)}{c^2 }
\]
For a mirror perfectly reflecting at low frequencies ($R_0 =1$), the mass
correction is negative. As the low temperature expression is valid for $T\ll
\hbar \omega _C $, one checks that the mass correction remains smaller than
the mirror's mass, using eq.(2) and noting that the delays are of the order
of the inverse of the cutoff frequency.

Viscous drag due to blackbody radiation has often been discussed,
particularly as a potential effect of the cosmic microwave background
radiation (see for instance ref. \cite{Friction15}). It follows from the
foregoing results that the background radiation may also affect the inertial
properties of scatterers.

\endmc

\end{document}